\documentclass[epj]{svjour}
\usepackage{graphicx,color}
\usepackage{amssymb}
\usepackage{amsmath}
\usepackage{latexsym}
\usepackage{amsxtra}
\usepackage{amscd}
\usepackage{dcolumn}
\usepackage{bm}
\usepackage[breaklinks]{hyperref}
\usepackage{url}
\usepackage{breakurl}
\usepackage{mathtools}
\usepackage{cuted}
\graphicspath{{./}}
\let\oldsqrt\sqrt
\def\sqrt{\mathpalette\DHLhksqrt}
\def\DHLhksqrt#1#2{\setbox0=\hbox{$#1\oldsqrt{#2\,}$}\dimen0=\ht0
\advance\dimen0-0.2\ht0
\setbox2=\hbox{\vrule height\ht0 depth -\dimen0}%
{\box0\lower0.4pt\box2}}

\newcommand{\nuc}[2]{$^{#1}$#2}

\newcommand{\bev}{$B(E2)$ value}
\newcommand{\bevs}{$B(E2)$ values}

\begin{document}

\title{Shape coexistence revealed in the \boldmath{$N=Z$} isotope \boldmath{\nuc{72}{Kr}} through inelastic scattering}
\author{K.~ Wimmer\inst{1,2,3}\thanks{\emph{e-mail: k.wimmer@csic.es}}\and        
  T.~Arici\inst{4,5}\and
  W.~Korten\inst{6}\and
  P.~Doornenbal\inst{2}\and
  J.-P.~Delaroche\inst{7}\and
  M.~Girod\inst{7}\and
  J.~Libert\inst{7}\and
  T.~R.~Rodr{\'i}guez\inst{8}\and
  P.~Aguilera\inst{9}\and
  A.~Algora\inst{10,11}\and
  T.~Ando\inst{1}\and
  H.~Baba\inst{2}\and
  B.~Blank\inst{12}\and
  A.~Boso\inst{13}\and
  S.~Chen\inst{2}\and
  A.~Corsi\inst{6}\and
  P.~Davies\inst{14}\and
  G.~de~Angelis\inst{15}\and
  G.~de~France\inst{16}\and
  D.~T.~Doherty\inst{6}\and
  J.~Gerl\inst{4}\and
  R.~Gernh\"{a}user\inst{17}\and
  T. Goigoux\inst{12}\and  
  D.~Jenkins\inst{14}\and
  G.~Kiss\inst{2,11}\and
  S.~Koyama\inst{1}\and
  T.~Motobayashi\inst{2}\and
  S.~Nagamine\inst{1}\and
  M.~Niikura\inst{1}\and
  S.~Nishimura\inst{2}\and
  A.~Obertelli\inst{6}\and
  D.~Lubos\inst{17}\and
  V.~H.~Phong\inst{2}\and
  B.~Rubio\inst{10}\and
  E.~Sahin\inst{18}\and
  T.~Y.~Saito\inst{1}\and
  H.~Sakurai\inst{1,2}\and
  L.~Sinclair\inst{14}\and
  D.~Steppenbeck\inst{2}\and
  R.~Taniuchi\inst{1}\and
  V.~Vaquero\inst{3}\and
  R.~Wadsworth\inst{14}\and
  J.~Wu\inst{2}\and
  M.~Zielinska\inst{6}
}

\institute{Department of Physics, The University of Tokyo, Hongo, Bunkyo-ku, Tokyo 113-0033, Japan\and
  RIKEN Nishina Center, 2-1 Hirosawa, Wako, Saitama 351-0198, Japan \and
  Instituto de Estructura de la Materia, CSIC, E-28006 Madrid, Spain \and
  GSI Helmholtzzentrum f\"{u}r Schwerionenforschung, D-64291 Darmstadt, Germany \and
  Justus-Liebig-Universit\"{a}t Giessen, D-35392 Giessen, Germany \and
  IRFU, CEA, Universit\'{e} Paris-Saclay, F-91191 Gif-sur-Yvette, France \and
  CEA, DAM, DIF 91297 Arpajon, France\and
  Departamento de F\'isica Te\'orica and Centro de Investigaci\'on Avanzada en F\'isica Fundamental, Universidad Aut{\'o}noma de Madrid, E-28049 Madrid, Spain\and
  Comisi\'{o}n Chilena de Energ\'{i}a Nuclear, Casilla 188-D, Santiago, Chile\and
  Instituto de Fisica Corpuscular, CSIC-Universidad de Valencia, E-46071 Valencia, Spain\and
  Institute for Nuclear Research (Atomki), H-4001 Debrecen, Hungary\and
  CENBG, CNRS/IN2P3, Universit\'{e} de Bordeaux F-33175 Gradignan, France\and
  Istituto Nazionale di Fisica Nucleare, Sezione di Padova, I-35131 Padova, Italy\and
  Department of Physics, University of York, YO10 5DD York, United Kingdom\and
  Istituto Nazionale di Fisica Nucleare, Laboratori Nazionali di Legnaro, I-35020 Legnaro, Italy\and
  GANIL, CEA/DSM-CNRS/IN2P3, F-14076 Caen Cedex 05, France\and
  Physik Department, Technische Universit\"{a}t M\"{u}nchen, D-85748 Garching, Germany\and
  Department of Physics, University of Oslo, PO Box 1048 Blindern, N-0316 Oslo, Norway
}

\date{\today}
\abstract{
  The $N=Z=36$ nucleus \nuc{72}{Kr} has been studied by inelastic scattering at intermediate energies. Two targets, \nuc{9}{Be} and \nuc{197}{Au}, were used to extract the nuclear deformation length, $\delta_\text{N}$, and the reduced $E2$ transition probability, $B(E2)$. The previously unknown non-yrast $2^+$ and $4^+$ states as well as a new candidate for the octupole $3^-$ state have been observed in the scattering on the Be target and placed in the level scheme based on $\gamma-\gamma$ coincidences. The second $2^+$ state was also observed in the scattering on the Au target and the $B(E2;\;2^+_2 \rightarrow 0^+_1)$ value could be determined for the first time. Analyzing the results in terms of a two-band mixing model shows clear evidence for a oblate-prolate shape coexistence and can be explained by a shape change from an oblate ground state to prolate deformed yrast band from the first $2^+$ state. This interpretation is corroborated by beyond mean field calculations using the Gogny D1S interaction.
  \PACS{
    24.50.+g \and	
    29.38.-c 	
  }
}
\maketitle

\section{Introduction}
Self conjugate $N=Z$ nuclei are of special interest for nuclear structure physics. Protons and neutrons occupy the same orbitals leading to a variety of features like enhanced proton-neutron pairing~\cite{frauendorf14}. In the region of $(N,Z) \approx 34,36$ shape transitions and coexistence are expected~\cite{moeller16}. Between the shell closures at $N=Z=28$ and $50$, the active orbitals are the proton and neutron $fp$ shell and the deformation-driving $g_{9/2}$ orbital. In the Nilsson diagram, deformed shell closures occur at 34, 36 on the oblate side, and 34, 38 on the prolate side. Experimentally, evidence for a shape transition has been obtained along the Kr isotopic chain: Low-energy Coulomb excitation experiments showed that the ground state of \nuc{78}{Kr} is prolate deformed~\cite{becker06}. In the more neutron-deficient Kr isotopes shape mixing occurs with equal mixing amplitudes for prolate and oblate configurations in the ground state of \nuc{74}{Kr}~\cite{clement07}. \nuc{72}{Kr} has been interpreted to have an oblate deformed ground state~\cite{bouchez03,gade05,briz15}.
Due to the dominant occurrence of prolate deformed ground states across the nuclear chart, the case of \nuc{72}{Kr} creates a rare opportunity to study this shape evolution as function of neutron number.

Spectroscopy of \nuc{72}{Kr} is known to high spin from fusion evaporation reactions~\cite{deangelis97,fisher01,kelsall01,andreoiu07}. In particular, the ground state band is well-deformed and expected to be prolate at high angular momenta. The discovery of a low-lying $0^+_2$ shape isomer by conversion-electron spectroscopy~\cite{bouchez03} proved the existence of shape coexistence in \nuc{72}{Kr}. A two-level mixing model also revealed evidence for an oblate dominated ground state.
The low-lying structure of \nuc{72}{Kr} was investigated using intermediate energy Coulomb excitation~\cite{gade05}. A $B(E2; 2^+_1\rightarrow 0^+_1)=999(129)$~e$^2$fm$^4$ results in a deformation parameter $\beta=0.33(2)$, which, when compared with self-consistent theoretical calculations, suggested an oblate ground-state deformation for \nuc{72}{Kr}. Later, a lifetime measurement found a slightly smaller $B(E2; 2^+_1\rightarrow 0^+_1)$ value of 810(150)~e$^2$fm$^4$, and additionally the $B(E2; 4^+_1\rightarrow 2^+_1) = 2720(550)$~e$^2$fm$^4$ value was determined~\cite{iwasaki14}. The large value for the $4^+_1\rightarrow 2^+_1$ transition, similar to the values for \nuc{74,76}{Kr}, suggested a transition to prolate deformation within the ground state band. Finally, the shape of the ground state of \nuc{72}{Kr} was also inferred from a measurements of its $\beta$ decay to \nuc{72}{Br}~\cite{briz15}. The comparison of the measured summed $B(\text{GT})$ distribution with QRPA calculations again supported an oblate shape for the ground state.

Various theoretical models predict shape coexistence in \nuc{72}{Kr}. Calculations using the finite-range liquid-drop model find triple shape coexistence in \nuc{72}{Kr}~\cite{moeller09}, with an oblate ground state minimum and a prolate minimum about 600 keV higher. An oblate ground state is also predicted by complex excited VAMPIR~\cite{petrovici02}, shell model Monte-Carlo~\cite{langanke03}, HFB-based beyond-mean-field models, using the Gogny D1S interaction and the 5DCH method~\cite{girod09,delaroche10} or the SCCM method~\cite{rodriguez14}, a schematic pairing plus quadrupole interaction with local QRPA~\cite{sato11}, the Skyrme SLy6 interaction~\cite{bender06} and the relativistic PC-PK1 interaction~\cite{fu13}. All calculations agree in their prediction for the $B(E2; 0^+_1\rightarrow 2^+_1)$ value with the experimental data. Some of these calculations predict a shape change along the yrast band, as suggested by a two-band mixing model~\cite{bouchez03}.
The present work supports this interpretation and provides new data on newly identified, non-yrast levels which allow to extend the two-band mixing model and the comparison with theoretical models.

\section{Experimental setup and analysis}\label{sec:ana}
The experiment was performed at the Radioactive Isotope Beam Factory, operated by the RIKEN Nishina Center and the Center for Nuclear Study, University of Tokyo. The \nuc{72}{Kr} beam was produced in projectile fragmentation of a 345~MeV/nucleon \nuc{78}{Kr} beam on a 5~mm thick primary Be target. The ions were separated and analyzed using the BigRIPS fragment separator~\cite{kubo12}. Unique identification of atomic number $Z$ and mass-to-charge ratio $A/q$ was achieved by measurements of time-of-flight, $B\rho$, and energy loss. \nuc{72}{Kr}, at an average intensity of 6000 particles per second and a purity of 64~\%, impinged on secondary Be and Au targets at an incident energy of 173.5~MeV/nucleon. The secondary target area was surrounded by the DALI2 NaI(Tl) array~\cite{takeuchi14} with 186 individual crystals. Energy and efficiency calibrations were performed using standard $\gamma$-ray calibration sources. The scattering angle at the secondary target was measured with two parallel plate avalanche counters (PPAC) located in front of the target and one behind the target with a precision of 5~mrad. Behind the target, reaction products were identified event-by-event in the ZeroDegree spectrometer using the same techniques as in BigRIPS.

\nuc{72}{Kr} has a low-lying isomeric $0^+$ state at 671~keV with a lifetime of $\tau = 38(3)$~ns~\cite{bouchez03}, which could be present in the beam. The flight time of the secondary beam from the production target to the secondary target amounted to $\sim450$~ns, much longer than the lifetime. However, since the decay from this first excited state is only possible by internal conversion and the secondary beam was fully stripped, the effective lifetime of the isomer was much longer and a significant fraction of the beam could be in an isomeric state. Therefore, the isomeric ratio was measured in a separate setting, where the beam was implanted into the WAS3ABi silicon detector array at the final focal plane of the ZeroDegree spectrometer~\cite{nishimura12}. 
The isomeric ratio was determined from the number of \nuc{72}{Kr} nuclei implanted in the first layer of WAS3ABi and the number of $0^+$ decays which were obtained by comparing the energy spectrum of the second layer with GEANT4~\cite{agostinelli03} simulations. The isomeric ratio obtained this way amounted to 4(1)~\% comparable to the isomeric ratio of 5.5(19)~\% in an experiment at lower energies at GANIL~\cite{bouchezphd}.

In order to extract the exclusive excitation cross section for states in \nuc{72}{Kr}, the measured $\gamma$-ray energy spectra were fitted with GEANT4~\cite{agostinelli03} simulations of the DALI2 response functions and a continuous background. After subtraction of indirect population, the $\gamma$-ray yield together with the number of detected outgoing ions, corrected for efficiency and transmission of the ZeroDegree spectrometer and the trigger efficiency, gave the total number of excitations. The cross section is calculated from the numbers of target ions and incident beam particles. The cross section values carry statistical as well as some systematic uncertainties. 
Several systematic uncertainties contribute to the total uncertainty for the excitation cross sections. The thickness of the target was measured by weighing the target, measuring the area and thickness, and also determined in the experiment from the energy loss of the beam. The remaining uncertainty on the number of target nuclei is thus small (less than 1~\%). The transmission of the ZeroDegree spectrometer as a function of scattering angle and momentum was investigated, and the measured $\gamma$-ray yield was corrected. Uncertainties on the transmission correction depend on the individual reaction setting, but for the inelastic scattering measurements presented here, the ZeroDegree spectrometer was centered on \nuc{72}{Kr} and thus \nuc{72}{Kr} lied fully within the momentum acceptance. In the measurement using the Au target, the scattering angle distribution is affected by the angular acceptance of ZeroDegree at large angles. This has been corrected, and systematic uncertainties of 1~\% were taken into account when calculating cross sections. Lastly, the efficiency of DALI2 was reproduced by the GEANT4 simulations within 5~\%, thus this systematic uncertainty had to be taken into account as well. All systematic uncertainties were added in quadrature.

The analysis follows the procedure described in Ref.~\cite{vaquero19}. In order to obtain the nuclear deformation length $\delta_\text{N}$ and the $B(E2)$ values from the measured cross sections, calculations with the coupled channels distorted waves code FRESCO~\cite{thompson88} were performed. In the present work a modified version including relativistic kinematics~\cite{moro18} was used. The optical model potentials were constructed following the approach described in Ref.~\cite{furumoto12}. The potentials were derived from the complex G-matrix interaction CEG07~\cite{furumoto09} using the microscopic folding model. The density distributions of protons and neutrons were based on the Sao Paulo parametrization~\cite{chamon02}. For the Be target the Sao Paulo density is not suitable because of the cluster structure of \nuc{9}{Be}. Alternatively, the \nuc{9}{Be} density was constructed in an $\alpha-\alpha-n$ cluster model~\cite{okabe79,hirabashi89}. For the Coulomb potential, a Coulomb radius $r_C=1.25$~fm was used. The final results for the \bevs\ change by about 1~\% if instead $r_C=1.20$~fm is used.
Calculations include both the excitation in the nuclear potential, determined by the nuclear deformation length $\delta_\text{N}$, and the excitation in the electromagnetic field of the target nucleus, depending on the $E2$ matrix element $\langle 0^+_\text{gs} || E2|| 2^+\rangle$. Both excitation modes interfere and the total excitation cross section thus depends on both amplitudes and their relative phase. The calculations for both targets, Be and Au, were thus performed in an iterative way. A first estimate for $\delta_\text{N}$ was obtained from the Be target data excluding any electromagnetic excitation, i.e. by setting $\langle 0^+_\text{gs} || E2|| 2^+\rangle=0$. In the next step, the value of $\delta_\text{N}$ was used in the calculation for the total excitation cross section for the scattering on the Au target. The resulting $E2$ matrix element was then used in the next iteration in the calculations for the Be target data and the procedure is repeated until convergence was reached. 

\section{Results}

\subsection{Inelastic scattering off a \nuc{9}{Be} target}\label{sec:nuclear}
\begin{figure}[h]
\centering
\includegraphics[width=\columnwidth]{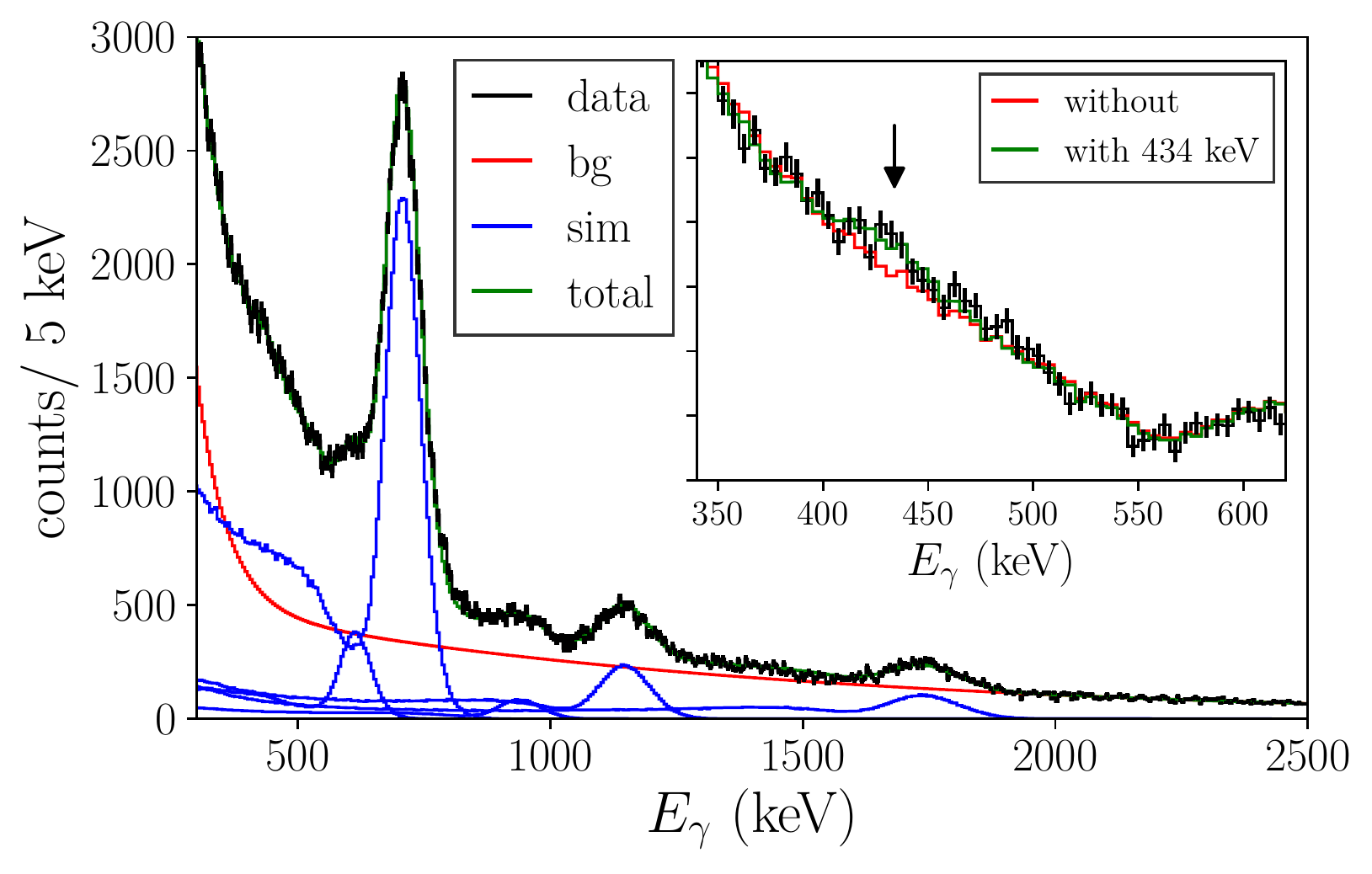}
\caption{Doppler-corrected $\gamma$-ray energy spectrum for the inelastic scattering of \nuc{72}{Kr} on a Be target. Only events where the multiplicity of DALI crystals with signals above threshold was less than five are shown. The data are superimposed with the result of a likely-hood fit (green) of the GEANT4 simulations of the DALI2 response function for the individual transitions (blue) and a continuous background (red). The insets shows a zoom on the low energy region. The fits were performed with and without an additional transition at 434~keV.}
\label{fig:bespectrum}
\end{figure}

\begin{figure}[h]
\centering
\includegraphics[width=\columnwidth]{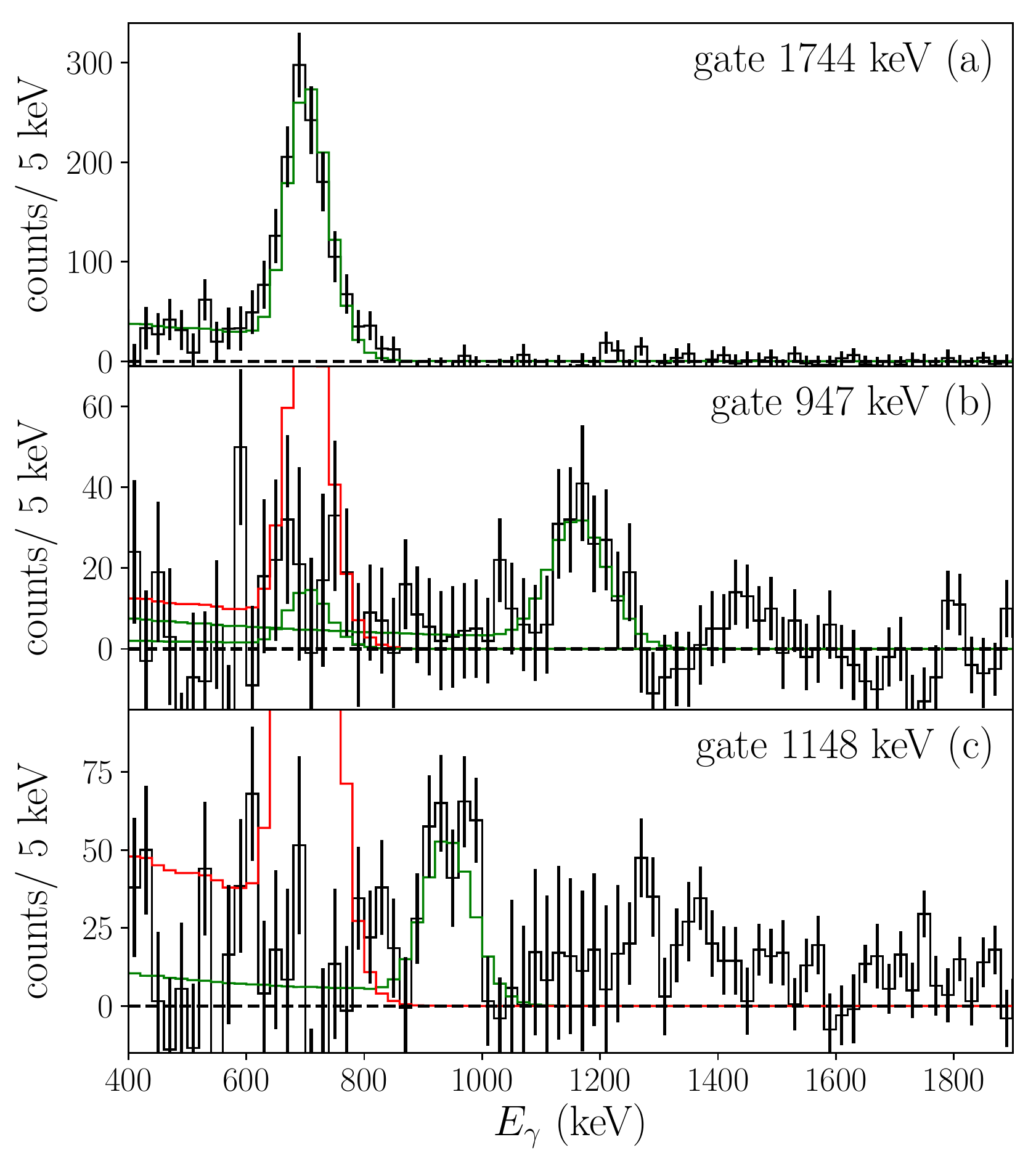}
\caption{$\gamma-\gamma$ coincidence spectra gated on the 1744, 947, and 1148~keV transitions. The green curves show simulated response functions scaled for the number of expected coincidences based on the proposed level scheme. For the 947 keV gate, this includes the branching ratio for the decay to the $2^+_1$ state determined from Fig.~\ref{fig:bespectrum}. For panels (b) and (c) additionally the expected coincidence yield for the 710~keV transition is shown (red curve), assuming that these transitions would feed the $2^+_1$ state directly.}
\label{fig:coinc} 
\end{figure}

\begin{figure}[h]
\centering
\includegraphics[width=0.8\columnwidth]{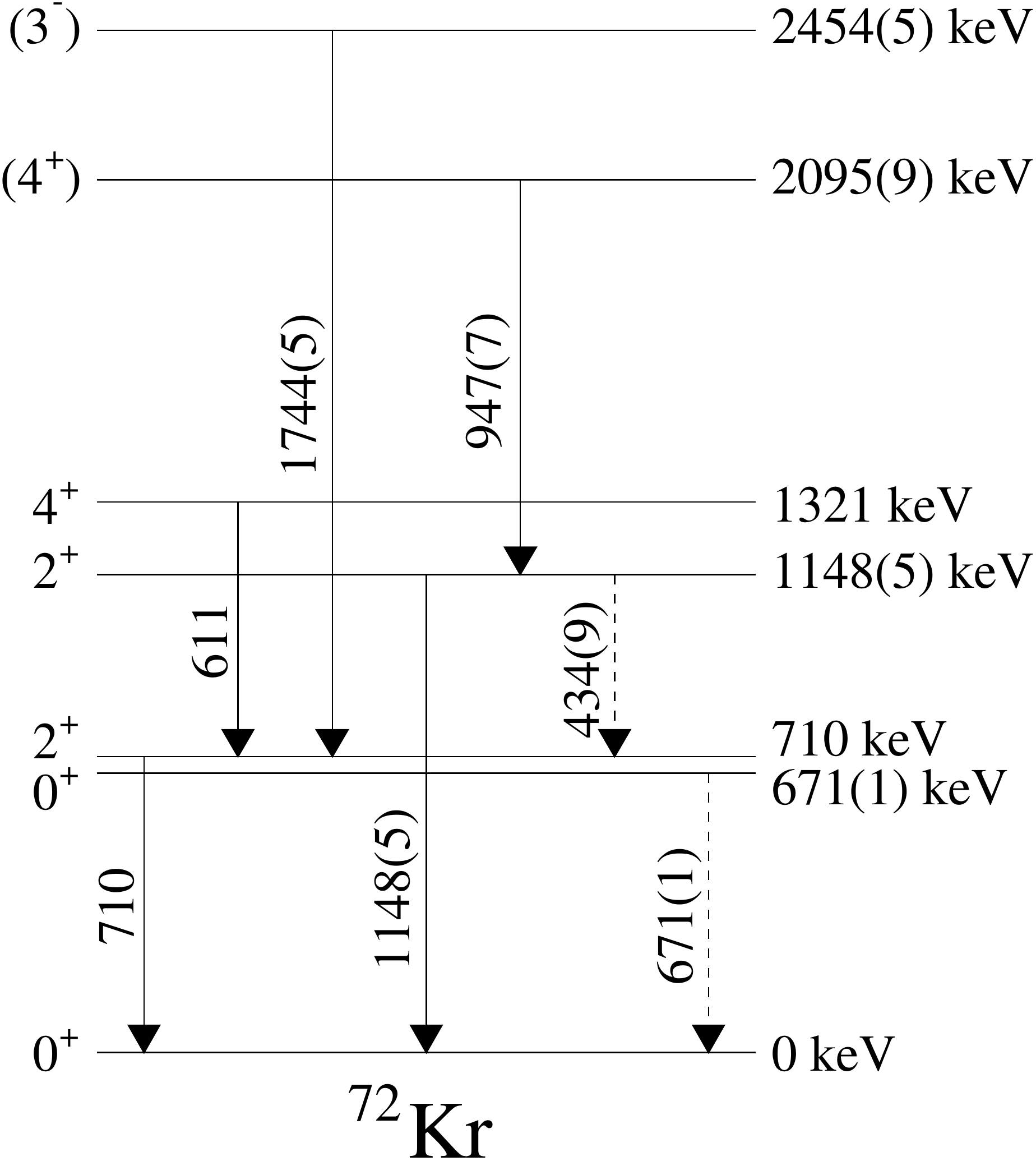}
\caption{Level scheme of \nuc{72}{Kr} deduced from this work. In addition to the transitions observed in the scattering of \nuc{72}{Kr} on the Be target, the excited $0^+_2$ state at 671(1)~keV~\cite{bouchez03} is included.}
\label{fig:level}
\end{figure}

The nuclear inelastic scattering was measured using a \nuc{9}{Be} reaction target. At the center of the 703(7)~mg/cm$^2$ thick target, the beam energy amounted to 146.6~MeV/nucleon. The $\gamma$-ray energy spectrum measured in coincidence with \nuc{72}{Kr} nuclei identified in the BigRIPS and ZeroDegree spectrometers is shown in Fig.~\ref{fig:bespectrum}.
Five transitions are observed in the spectrum. The 710 and 611~keV transitions were known before and assigned to the $2^+_1 \rightarrow 0^+_\text{gs}$ and $4^+_1 \rightarrow 2^+_1$ decays, respectively. The transitions at 947(7), 1148(5), and 1744(5)~keV were observed for the first time. The transition energies were determined using a maximum likely-hood fit of simulated response functions. Known transition energies in \nuc{72}{Kr} and neighboring isotopes were reproduced within 5~keV. The level scheme was constructed using the information from $\gamma$ coincidences shown in Fig.~\ref{fig:coinc}.
The 1744~keV transition was in coincidence with the $2^+\rightarrow 0^+_\text{gs}$ transition, establishing a new state at 2454~keV. No direct ground state transition was observed. This state is therefore a fitting candidate for a $3^-$ state, which is typically strongly excited in inelastic nuclear scattering~\cite{kibedi02}. For example, an octupole deformation of $\beta_3=0.1$ gives rise to a cross section $\sigma(3^-)$ of about $2-3$~mb.
The 947 and 1148~keV transitions are in mutual coincidence (see Fig.~\ref{fig:coinc} (b,c)), but not in coincidence with the 710~keV transition. As the 1148~keV transition has higher intensity, it is placed below, feeding the ground state. Based on the intensity pattern, systematics of neighboring isotopes, and isotones they are assigned as the $2^+_2$ and $(4^+_2)$ states at 1148(5) and 2095(9)~keV, respectively. However, these two transitions could in principle also be built on top of the isomeric $0^+_2$ state at 671~keV~\cite{bouchez03} instead of the ground state. In order to exclude this possibility we looked for other possible decay branches of the second $2^+$ state.
A ground state transition from a state at 1819~keV can be excluded ($<1$~\% branching ratio at 95~\% confidence level). Assuming a $2^+_2$ state at 1148(5) keV, a decay to either the 710 keV $2^+_1$ or the 671 keV $0^+_2$ state would allow for transitions at 438(5) or 477(5)~keV. The inset of Fig.~\ref{fig:bespectrum} shows the low energy region of the $\gamma$-ray energy spectrum. An excess of counts around these energies can be seen. An additional transition was therefore included in the fit. By varying the simulated transition energy, a value of 434(9)~keV was obtained from a maximum likely-hood fit, in excellent agreement with the difference between the $2^+_2$ and $2^+_1$ states. The branching ratio amounts to 16(3)~\% of the ground state transition. For the 477~keV $2^+_2\rightarrow 0^+_2$ transition, an upper limit of 3.5~\% (at 95~\% confidence level) of the 1148 keV transition was determined.
The proposed level scheme is shown in Fig.~\ref{fig:level}.
A transition at 1139~keV, close to the proposed $2^+_2 \rightarrow 0^+_1$ transition, has been previously observed~\cite{kelsall01}. It was assigned to the decay of a $J^\pi = (3^-)$ state at 1849~keV, but it is not clearly separated from a close-lying 1134~keV transition. The absence of a coincidence with the $2^+_1 \rightarrow 0^+_1$ transition established in this work shows that a different state is observed here. A transition at around 1150~keV has also been observed in the two-neutron removal reaction \nuc{9}{Be}(\nuc{74}{Kr},X)\nuc{72}{Kr}~\cite{iwasaki14}. The lifetime of the decaying state, $\tau = 2^{+1}_{-0.5}$~ps, was determined from a line-shape analysis. Using this lifetime and the branching ratio of the present work the reduced transition probability amounts to $B(E2;\;2^+_2 \rightarrow 0^+_1) = 176 ^{+67}_{-59}$~e$^2$fm$^4$.

The results for the exclusive cross sections for the population of the $2^+$ states are summarized in Table~\ref{tab:cs}.
\begin{table}[h]
  \caption{Cross sections for the excitation of the $2^+$ states on the Be and Au targets. The numbers are inclusive with respect to the beam composition of ground and isomeric state, i.e., $\sigma = (1-r)\sigma(0^+_1 \rightarrow 2^+) + r \sigma(0^+_2 \rightarrow 2^+)$ with the isomeric ratio $r = 4(1)$~\%. Deformation parameters and reduced transition probabilities for the excitation of the $2^+$ states in \nuc{72}{Kr} are given in the lower part of the table. The numbers in parenthesis represent the statistical, systematical, and theoretical uncertainties, respectively. See text for details.}
  \centering
  \begin{tabular}{lcc}
    & $2^+_1$ & $2^+_2$ \\ 
    \hline\noalign{\smallskip}
    &\multicolumn{2}{c}{inelastic scattering off Be target}\\
    $\sigma$ (mb) & 27.2(4)(15) & 4.5(3)(2)\\
    &\multicolumn{2}{c}{inelastic scattering off Au target}\\
    $\sigma$ (mb) & 468(9)(29) & 79(4)(5)\\
    \noalign{\smallskip}\hline\noalign{\smallskip}
    & $0^+_1 \rightarrow 2^+_1$ & $0^+_1 \rightarrow 2^+_2$ \\     
    $\delta_\text{N}$ (fm) & 1.541(11)(46)(38) & 0.613(21)(25)(16) \\
    $\beta_\text{N}$ & 0.309(2)(9)(8) & 0.123(4)(5)(3)\\
    $B(E2)$ e$^2$fm$^4$ & 4023(81)(290)(380) & 665(39)(58)(63)\\ 
    $\beta_\text{C}$ & 0.296(3)(11)(13) & 0.112(3)(4)(5)\\
    \noalign{\smallskip}\hline
  \end{tabular}
  \label{tab:cs}
\vspace{-0.2cm}
\end{table}
The nuclear deformation length was extracted from the measured cross section on the Be target as discussed in Section~\ref{sec:ana}. The excitation of the $2^+$ states by the Coulomb field of the Be target was taken into account and the $E2$ matrix elements determined from the scattering on the Au target described in Section~\ref{sec:coulex} were used.  In this way a nuclear deformation length of $\delta_\text{N} = 1.54(6)$~fm and $0.61(4)$~fm was obtained for the $2^+_1$ and $2^+_2$ states, respectively. These values correspond to deformation parameters $\beta_\text{N} = \delta_\text{N}/R = 0.31(1)$ and $0.12(1)$ with the nuclear radius $R=1.2\cdot A^{1/3}$~fm. If instead of the $\alpha-\alpha-n$ cluster model, the Sao Paulo parametrization is used for the Be optical model potential, the cross sections change by less than 2~\%.
The isomeric ratio of the beam has little influence on the extracted deformation length for the $2^+_1$ state. For the second $2^+$ state, the uncertainty amounts to 5~\%.
The values for the deformation parameters for the excitation from the ground state are listed in Table~\ref{tab:cs}.

\subsection{Coulomb excitation on a \nuc{197}{Au} target}\label{sec:coulex}
\begin{figure}[h]
\centering
\includegraphics[width=\columnwidth]{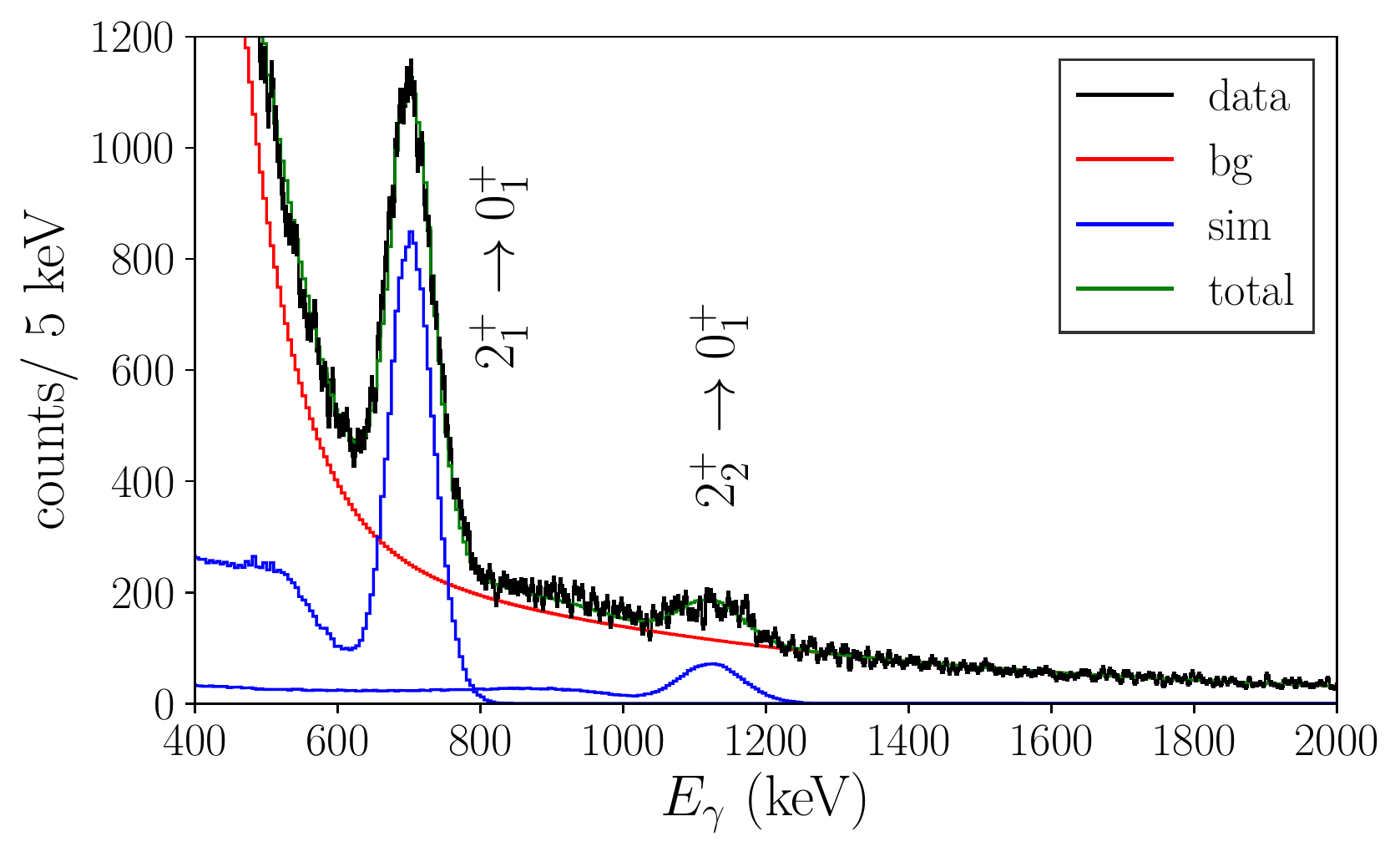}
\caption{Doppler corrected $\gamma$-ray energy spectrum for the inelastic scattering of \nuc{72}{Kr} on a \nuc{197}{Au} target. The Doppler correction assumes $\gamma$-ray emission at the velocity in the center of the target. The data are fitted with simulated response functions for the two transitions and a continuous background (red). Only forward DALI2 crystals ($\theta_\text{lab}<75^\circ$) are shown to reduce background from atomic processes.}
\label{fig:auspectrum}
\end{figure}

\begin{figure}[h]
  \centering
\includegraphics[width=\columnwidth]{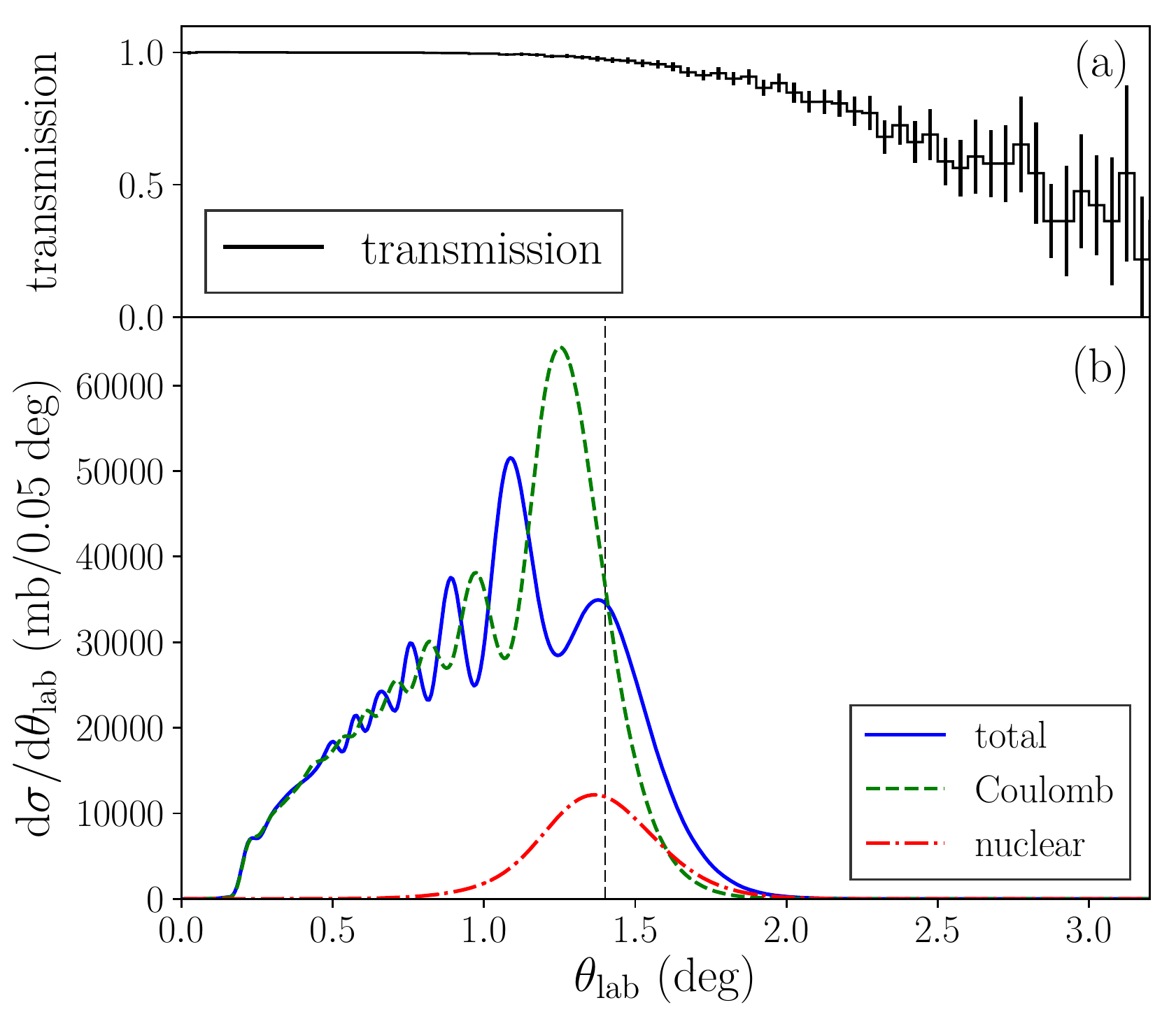}
\caption{(a) Transmission of the ZeroDegree spectrometer as a function of the scattering angle $\theta_\text{lab}$ for the scattering on the Au target. (b) Theoretical differential cross sections of Coulomb, nuclear and total excitation cross section to the first $2^+$ state. The dashed line indicates the scattering angle corresponding to the impact parameter for a ``safe'' distance.}
\label{fig:kr72_theoang}
\end{figure}

The Doppler corrected $\gamma$-ray energy spectrum for the measurement with the 405(4)~mg/cm$^2$ thick Au target is shown in Fig.~\ref{fig:auspectrum}. The Doppler correction is applied for a mid-target energy of 163.8~MeV/nucleon. 
Two transitions are observed, the known 710~keV $2^+_1 \rightarrow 0^+_1$ transition and the 1148~keV transition that was newly assigned to the $2^+_2 \rightarrow 0^+_1$ decay (see Section~\ref{sec:nuclear}). The $\gamma$-ray yield was extracted by fitting simulated response functions for the two transitions and a double exponential background to the data. The resulting fit is shown in Fig. ~\ref{fig:auspectrum}. For the angular distribution prolate alignment was assumed. Using the semi-classical Alder-Winther theory of Coulomb excitation~\cite{winther79}, 96~\% (92~\%) prolate alignment is predicted for the $2^+_1$ ($2^+_2$) state. The $\gamma$-ray yield was determined using only forward DALI2 crystals ($\theta_\text{lab}<75^\circ$) to reduce background from atomic processes, but the result is consistent with other angular ranges for the $\gamma$ rays if prolate alignment is assumed.
In order to determine the cross section, the $\gamma$-ray yield has to be corrected for the finite transmission of the ZeroDegree spectrometer shown in Fig.~\ref{fig:kr72_theoang} (a).
In the region of scattering angles below 2$^\circ$, where the maximum yield is expected, the transmission is well determined and corrections are small. Using this angle dependent transmission, the exclusive cross sections for the excitation of the two $2^+$ states have been determined. For the $2^+_1$ state, indirect population from the $2^+_2$ state with the branching ratio determined in Section~\ref{sec:nuclear} was subtracted. No other discrete transition feeding the $2^+$ states was observed in the present experiment. No excess of counts beyond the background was observed, limiting this indirect population to 2~\% for any individual transition. However, the strength might be distributed over several states, and thus not directly observed. Theoretical calculations predict the $2^+_3$ and higher states to be significantly less collective than the $2^+_1$ state. In the less neutron-deficient \nuc{74,76}{Kr} isotopes the $B(E2;\;0^+_1 \rightarrow 2^+_3)$ values are factors of 20 and 50 smaller than for the $2^+_1$ state~\cite{clement07}.
In order to estimate the contributions to the indirect population of the $2^+_{1,2}$ states, theoretical calculation were employed (see Section~\ref{sec:hfb}). The cross sections for all states up to the proton separation energy were calculated using the predicted $B(E2)$ values, and predicted decay branching ratios to the $2^+_{1,2}$ states were assumed. The indirect population of the $2^+_1$ state was dominated by the $2^+_{2,3}$ states, states above 3~MeV excitation energy could be ignored in the estimate. The total theoretical indirect population amounts to 6.9~mb for the $2^+_1$ state and 2.8~mb for the $2^+_2$ state. This reduces the cross section for the $2^+_1$ state by less than 2~\%, and by 4~\% for the $2^+_2$ state. Systematic uncertainties for the cross sections are estimated to be of the same order.

\begin{figure}[h]
\includegraphics[width=\columnwidth]{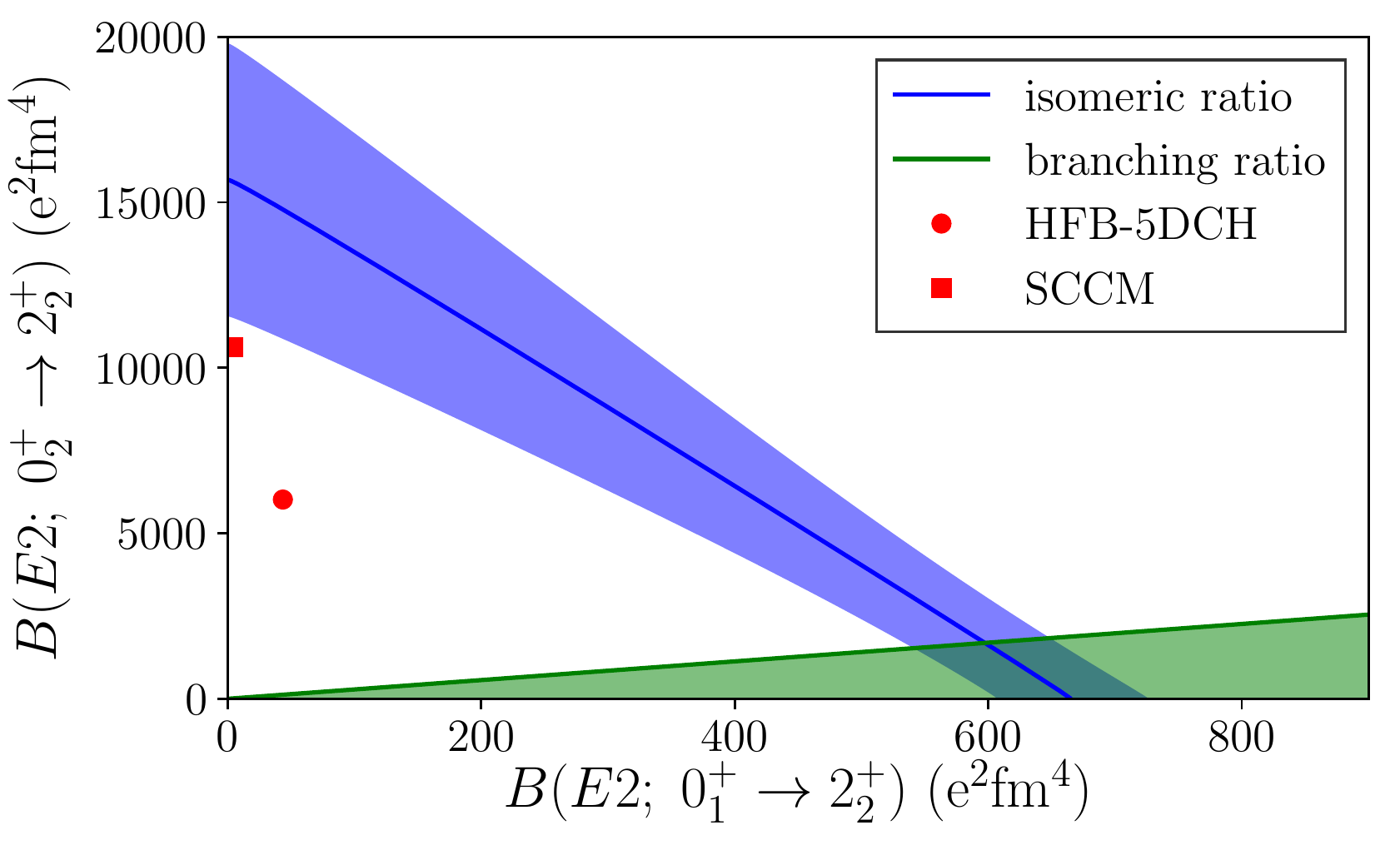}
\caption{Constraining the \bevs\ for the second $2^+$ state. The blue band indicates the values of $B(E2; 0^+_x\rightarrow 2^+_2)$ that agree with the experimental value of the cross section observed for the $2^+_2$ state given the measured isomeric ratio of the beam. The limit on the branching ratio for the $2^+_2 \rightarrow 0^+_2$ of 5~\% of the ground state transition results in a limit for the \bevs\ shown in green. The red dot indicates the calculated \bevs\ (see Section~\ref{sec:hfb}).}
\label{fig:kr72_second}
\end{figure}

The isomeric content of the beam has two influences on the extracted \bev. Firstly, the amount of beam particles that is in an isomeric state when arriving at the secondary reaction target needs to be subtracted from the \nuc{72}{Kr} beam intensity to determine the $B(E2; 0^+_1\rightarrow 2^+_x)$ values. Secondly, the $2^+$ states can also be excited from the $0^+_2$ isomeric state, and therefore lead to a higher or lower yield of $\gamma$-ray counts compared to the absence of an isomeric state in the beam.
For the first $2^+$ state the isomeric content of the beam presents a small correction of 2~\%, with a systematic uncertainty of the same order. 
In the case of the second $2^+$ state, the theoretical \bevs, together with the experimental isomeric ratio in the beam yield $\sigma(2^+_2) = 38$~mb for the HFB-5DCH calculations, compatible with the experimental data. The SCCM calculations yield a cross section of 99~mb for the $2^+_2$ state.
But, as discussed below, the calculations are at variance with the upper limit for the branching ratio of the decay of the $2^+_2$ state. Fig.~\ref{fig:kr72_second} shows the experimental constraints on the \bevs\ that reproduce the measured cross section.
The cross section for the population of the $2^+_2$ state together with the measured isomeric ratio results in possible $B(E2; 0^+_x\rightarrow 2^+_2)$ values shown by the blue band in Fig.~\ref{fig:kr72_second}. The upper limit for the branching ratio of the $2^+_2 \rightarrow 0^+_2$ transition of 3.5~\% limits the range of possible \bevs\ as shown by the green band and allows for the determination of an upper limit for the $B(E2; 0^+_2\rightarrow 2^+_2)$ value of  $< 1833$~e$^2$fm$^4$. 

The exclusive cross sections listed in Table~\ref{tab:cs} include both the excitation through the nuclear and the Coulomb interaction (see Fig.~\ref{fig:kr72_theoang} (b)). The nuclear interaction only contributes for sufficiently small impact parameters, corresponding to a maximum scattering angle for which the excitation can be considered to be purely electromagnetic. The ``safe'' distance, typically the sum of two radii with additional 2~fm, corresponds to a scattering angle of 1.4$^\circ$ in the present case. As shown in Fig.~\ref{fig:kr72_theoang} (b) the nuclear interaction already contributes to the excitation cross section at this angle. Elimination of the nuclear excitation would thus require a gate on scattering angles less than 1$^\circ$ which reduces the statistics by a factor of two. Instead distorted wave coupled channels calculations were used, taking into account the nuclear and electromagnetic excitations. Fig.~\ref{fig:kr72_theoang} (b) shows the differential cross sections for the combined nuclear and Coulomb excitation of \nuc{72}{Kr} on \nuc{197}{Au}. The interference of the two excitation modes to the total excitation cross section can be observed in the reduction of the maximum of the dominating Coulomb excitation process at 1.2$^\circ$. The experimental resolution for the scattering angle is not sufficient to resolve the structure of the differential cross section. In order to determine the $B(E2)$ value, the nuclear deformation length extracted in Section~\ref{sec:nuclear} was fixed in the FRESCO calculation, while the $E2$ matrix element was varied to reproduce the experimental cross section. The statistical error includes, besides the number of beam particles, contributions from the fitting of the spectrum shown in Fig.~\ref{fig:auspectrum} and from the subtraction of the observed feeding from the $2^+_2$ state. The systematical uncertainties comprise the uncertainty of the target thickness, the beam intensity, the efficiency of the DALI2 array, and the transmission through the ZeroDegree spectrometer. Also the effects of the indirect population estimated above and the isomeric content of the beam were included in the systematic uncertainty. The individual contributions to the experimental error are shown in Table~\ref{tab:error}.
\begin{table}[h]
\vspace{-0.2cm}
  \caption{Contribution to the error in the extracted \bev\ for the two $2^+$ states in \nuc{72}{Kr}. See text for details.}
  \centering
  \begin{tabular}{lcc}
    & $2^+_1$ & $2^+_2$ \\ 
    \noalign{\smallskip}\hline
    \multicolumn{3}{c}{statistical for $\sigma(2^+)$}\\
    fitting of spectrum               & 1.3~\% & 6.5~\%\\
    subtraction of observed feeding   & 1.4~\% & -     \\
    \noalign{\smallskip}\hline\noalign{\smallskip}
    \multicolumn{3}{c}{systematic for $\sigma(2^+)$}\\
    particle gates                    & \multicolumn{2}{c}{1.2~\%}\\
    isomeric content of the beam      & 2.0~\% & 5.0~\%\\
    unobserved feeding                & 1.5~\% & 4.0~\%\\
    ZeroDegree efficiency             & \multicolumn{2}{c}{0.2~\%}\\
    ZeroDegree transmission           & \multicolumn{2}{c}{1.0~\%}\\
    DALI2 efficiency                  & \multicolumn{2}{c}{5.0~\%}\\
    target thickness                  & \multicolumn{2}{c}{1.0~\%}\\
    trigger efficiency                & \multicolumn{2}{c}{2.0~\%}\\
    $\gamma$-ray angular distribution & \multicolumn{2}{c}{2.0~\%}\\
    \noalign{\smallskip}\hline\noalign{\smallskip}
    \multicolumn{3}{c}{theoretical for $B(E2; 0^+_1\rightarrow 2^+_x)$}\\
    reaction energy                   & \multicolumn{2}{c}{$<0.2$~\%}\\
    optical potentials                & \multicolumn{2}{c}{$8$~\%}\\
    relativistic dynamics             & \multicolumn{2}{c}{5~\%}\\
    nuclear deformation               & 0.7~\%& 1.0 ~\%\\    
    \noalign{\smallskip}\hline
  \end{tabular}
  \label{tab:error}
\vspace{-0.2cm}
\end{table}
In addition to the experimental statistical and systematic uncertainties, an uncertainty arising from the reaction model contributes.
For the analysis, the mid-target energy was used in the calculation of the optical model potential and the reaction cross section. However, the beam lost about 11~\% of its initial energy in the Au target. The effect of the beam energy on the excitation cross section was investigated and found to be very small ($<0.2$~\%). 

The present analysis used a modified version of the FRESCO code which includes relativistic kinematics~\cite{moro18}. Dynamical relativistic corrections to the potentials were not performed. However, these are considered to be small for the present projectile energies~\cite{moro18,vaquero19} and therefore a conservative 5~\% theoretical uncertainty due to the incomplete description of the reaction dynamics has been employed.

The reaction model calculations used optical model potentials based on the Sao-Paulo neutron and proton density distributions, which were adjusted for stable nuclei. In the present case of very neutron-deficient nuclei, this parametrization might not be valid anymore. Thus, also optical model potentials were constructed using the $t\rho\rho$-approximation~\cite{hussein91} with different Skyrme interactions to calculate the densities. These lead to slightly different values for the \bev, and thus a theoretical uncertainty of 8~\% has been assumed for the effect of the optical model potential.

Lastly, the uncertainty for the nuclear deformation extracted in Section~\ref{sec:nuclear} influences the extracted \bev. For both states the nuclear deformation length $\delta_\text{N}$ has been varied within the error bars to estimate the contribution to the theoretical uncertainty for the \bev. All uncertainties are listed in Table~\ref{tab:error} for both states, and are added in quadrature to obtain the final uncertainties for the \bevs\ and the deformation parameters $\beta_\text{C}$ shown in Table~\ref{tab:cs}.

For the first $2^+$ state, $B(E2; 0^+_1\rightarrow 2^+_1)\allowbreak=\allowbreak4023(81)_\text{stat.}\allowbreak(290)_\text{syst.}\allowbreak(380)_\text{theo.}$~e$^2$fm$^4$ was obtained. The  value is in good agreement with the result of a lifetime measurement, ($B(E2; 0^+_1\rightarrow 2^+_1)=4050(750)$~e$^2$fm$^4$)~\cite{iwasaki14} and slightly lower than  $B(E2)= 4997(647)$~e$^2$fm$^4$ determined in an earlier Coulomb excitation measurement~\cite{gade05}. 
The difference is partially related to the fact that in the earlier measurement the feeding from the second $2^+$ state could not be taken into account. 
The value for the $B(E2; {0^+_1\rightarrow 2^+_2})\allowbreak=\allowbreak 665(39)_\text{stat}(58)_\text{syst.}(63)_\text{theo.}$~e$^2$fm$^4$ was determined for the first time in this work. For the $2^+_2\rightarrow 2^+_1$ transition the measured branching ratio also allows for the extraction of the $B(E2; 2^+_1\rightarrow 2^+_2)\allowbreak=\allowbreak13800(2600)_\text{stat}\allowbreak(1200)_\text{syst.}\allowbreak(1200)_\text{theo.}$~e$^2$fm$^4$, under the assumption of a pure $E2$ transition with a mixing ratio $\delta(M1/E2)$ $=0$. The branching ratio limit for the decay to the $0^+_2$ state provides an upper limit $B(E2; 0^+_2\rightarrow 2^+_2)< 1880$~e$^2$fm$^4$.

\begin{figure}[h]
  \centering
\includegraphics[width=\columnwidth]{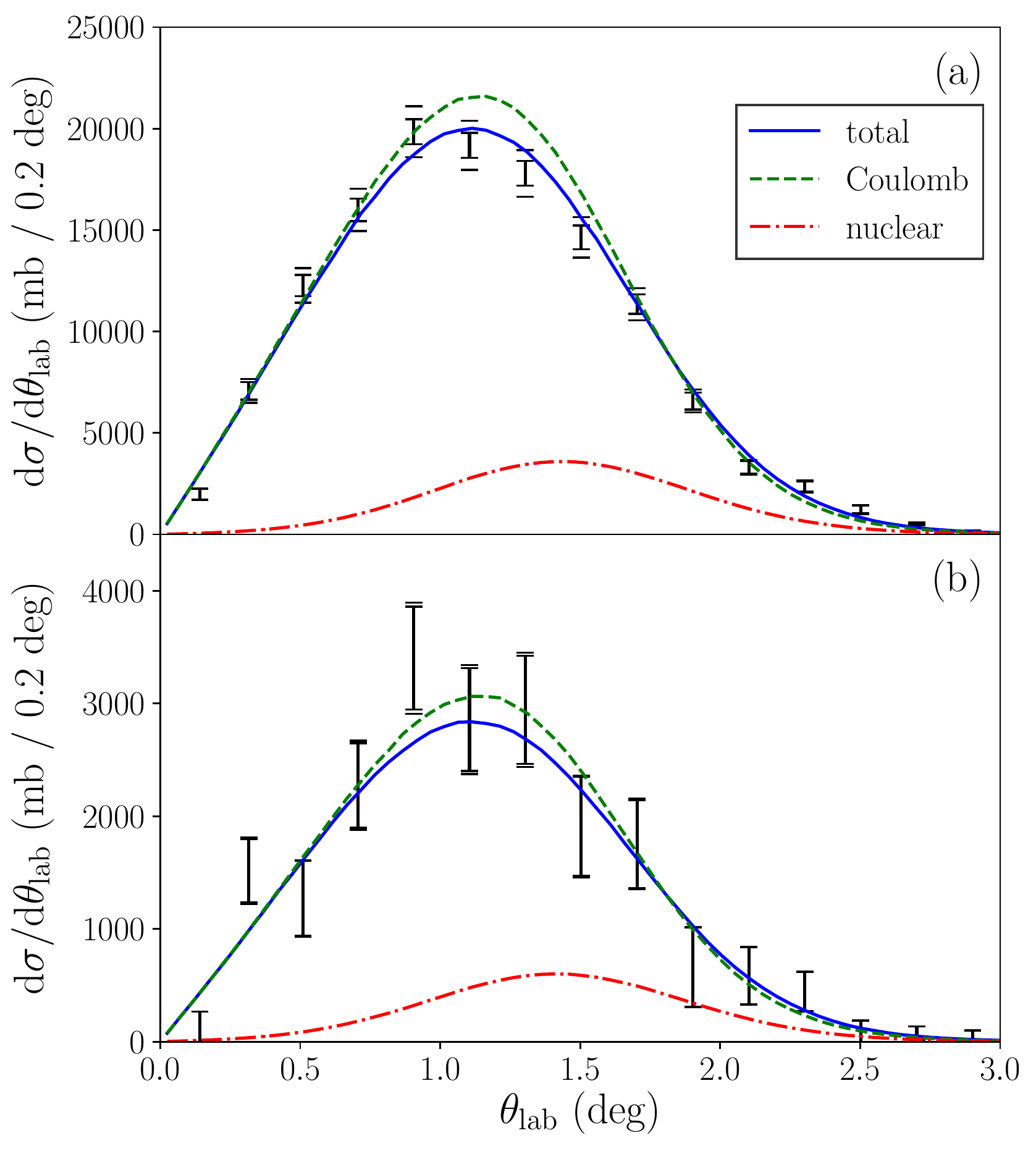}
\caption{Differential excitation cross section for the $2^+_1$ state (a) and the $2^+_2$ state (b). Panel (a) shows the same calculations as Fig.~\ref{fig:kr72_theoang} (b), but convoluted with the experimental resolution. The error bars are showing the statistical uncertainties and the total uncertainties with the systematic contribution added in quadrature. The data are compared with the results of the coupled channels calculations (blue line). The contribution of the Coulomb excitation alone (green line) is reduced by the interference with the nuclear excitation (red line).}
\label{fig:kr72_expang}
\end{figure}

In Fig.~\ref{fig:kr72_expang}, the calculated differential cross sections are compared to the experimental data.
The experimental resolution for the scattering angle amounts to 0.43$^\circ$ and is dominated by the angular straggling in the 0.2~mm thick Au target. Therefore, the theoretical differential cross sections were convoluted with the experimental resolution, including the angular straggling in the target and the position resolution of the PPAC detectors.
The shape of the calculated differential cross section agrees very well for the known $2^+_1$ state with the experimental data. In particular, the destructive interference of Coulomb and nuclear excitation is visible in the reduction of the cross section near the maximum of the distribution.
Also for the $2^+_2$ state good agreement is observed, providing additional confirmation for the total angular momentum for this state as the shape of the angular distribution would be narrower(wider) for $1^+$($3^-$) possibilities.
The present experiment establishes the second $2^+$ state at 1148~keV in \nuc{72}{Kr}. 

\section{Discussion}
The \bev\ can be related to the deformation parameter $\beta_\text{C}$ by~\cite{raman01}:
\begin{equation}
  B(E2;\;0^+_1 \rightarrow 2^+_1)=\left(\frac{3}{4\pi}ZeR^2\beta_\text{C}\right)^2
  \end{equation}
with the nuclear radius $R=1.2\cdot A^{1/3}$~fm. The $\beta_\text{C}$ values together with their uncertainties are listed in Table~\ref{tab:cs}. The results with two very different deformations indicate shape coexistence in \nuc{72}{Kr}.
The values for the deformation $\beta_\text{N/C}$ extracted from the two data sets agree very well within their uncertainties. 

\subsection{Two-band mixing model}\label{sec:mixing}
\begin{figure}[h]
\centering
\includegraphics[width=0.8\columnwidth]{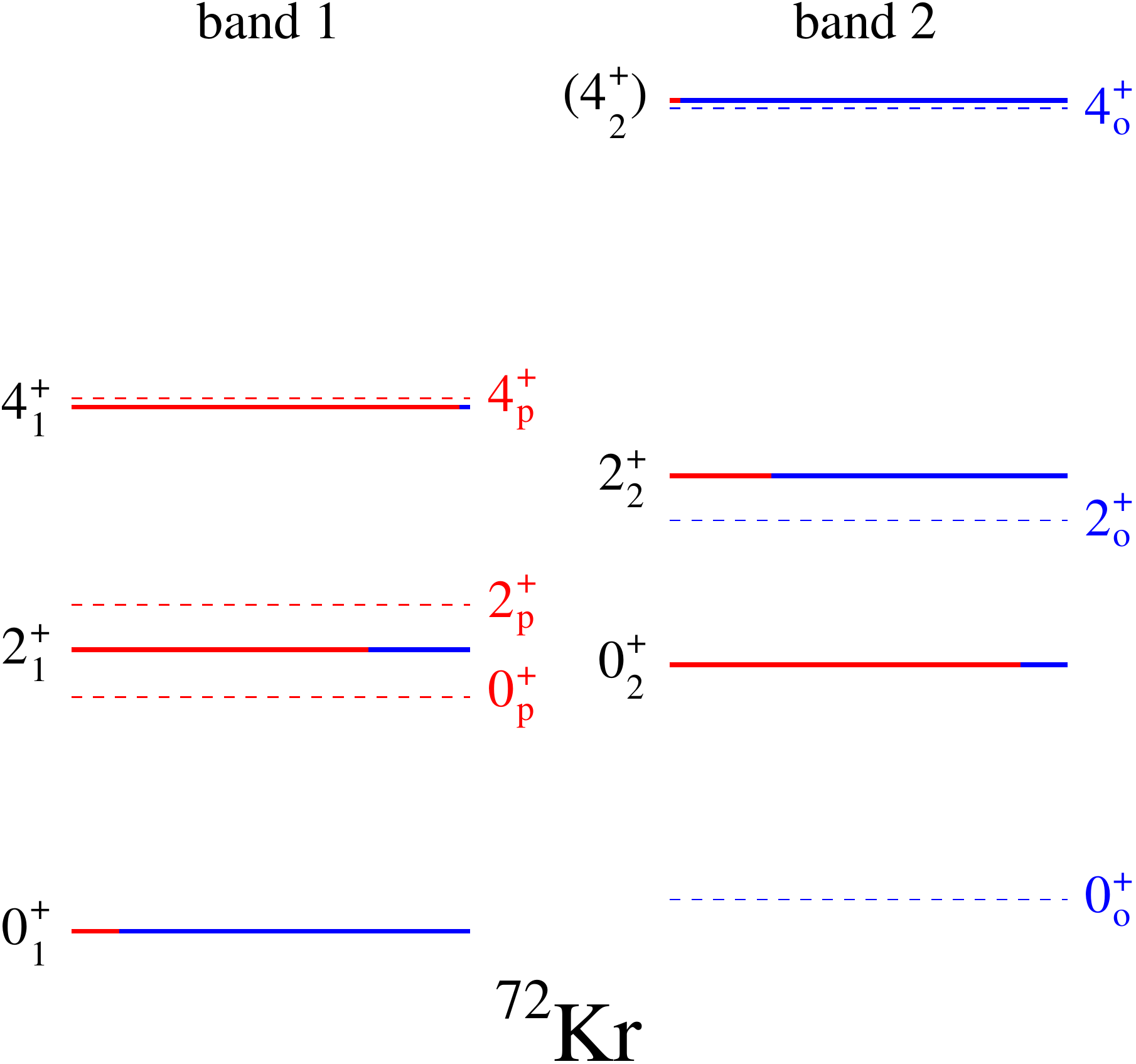}
\caption{Two-band mixing model for \nuc{72}{Kr}. The unperturbed energies of prolate states are extrapolated from higher spins following the method described in~\cite{piercey81,becker99,bouchez03}. The length of the colored bars for the physical $J^+_i$ states shows the amplitudes of the prolate (red) and oblate (blue) configurations in the wave functions.}
\label{fig:bandmix}
\end{figure}

Following earlier work for the proton-rich Kr nuclei~\cite{piercey81,becker99,bouchez03}, the mixing of states can be investigated by a simple two-band mixing model. Here it is assumed that the physical $J^\pi = 0^+, 2^+,$ and $4^+$ states are the results of the mixing of two pure configurations
\begin{eqnarray}
  |J^+_1\rangle &=& +a_J|J^+_\text{p}\rangle +b_J|J^+_\text{o}\rangle \nonumber\\
  |J^+_2\rangle &=& -b_J|J^+_\text{p}\rangle +a_J|J^+_\text{o}\rangle \label{eq:wavefunctions}
\end{eqnarray}
where $a$ and $b$ denote the amplitudes of the prolate (p) and oblate (o) configurations with $a_J^2+b_J^2=1$. From the energy differences of these states, $\Delta E_\text{per} = E_2(J) - E_1(J)$ and $\Delta E_\text{unp} = E_o(J) - E_p(J)$, the mixing matrix element $V$ and the mixing amplitudes $a$ and $b$ can be determined. At high spins $J\geq 6$ the yrast bands of the proton-rich Kr isotopes are well deformed and assumed to be prolate~\cite{piercey81,fisher03}. In order to determine the energies of the unperturbed states, the moments of inertia in the yrast band have been extrapolated to $J^\pi = 0^+$ using a variable moment of inertia parametrization $I=I_0+\omega^2I_1$. As demonstrated in Ref.~\cite{bouchez03}, this leads in \nuc{72}{Kr} to an inversion of the two shapes, with the excited $0^+_2$ as the prolate band head, and the amplitude of the oblate configuration in the ground state amounts to $b_0^2=0.881$. Considering only the known $2^+$ and $4^+$ states and ignoring the potential mixing with other states, the same procedure can now be applied to higher spins.  
The results of the two-band mixing model for the admixture in the $0^+$, $2^+$, and $4^+$ states in \nuc{72}{Kr} are also shown in Fig.~\ref{fig:bandmix}.
For \nuc{72}{Kr}, the oblate amplitude in the $2^+_1$ state is already significantly smaller with $b_2^2=0.256$. 
This shows that, while the ground state has a large oblate fraction, a quick evolution towards prolate character is observed in the ground state band. This is in agreement with the interpretation of the large $B(E2; 4^+_1\rightarrow 2^+_1)$ value which suggests a similar structure of the $2^+_1$ and $4^+_1$ states~\cite{iwasaki14}. The $4^+_1$ state is almost purely prolate with $a_4^2 = 0.972$. 

\begin{figure}[h]
\centering
\includegraphics[width=\columnwidth]{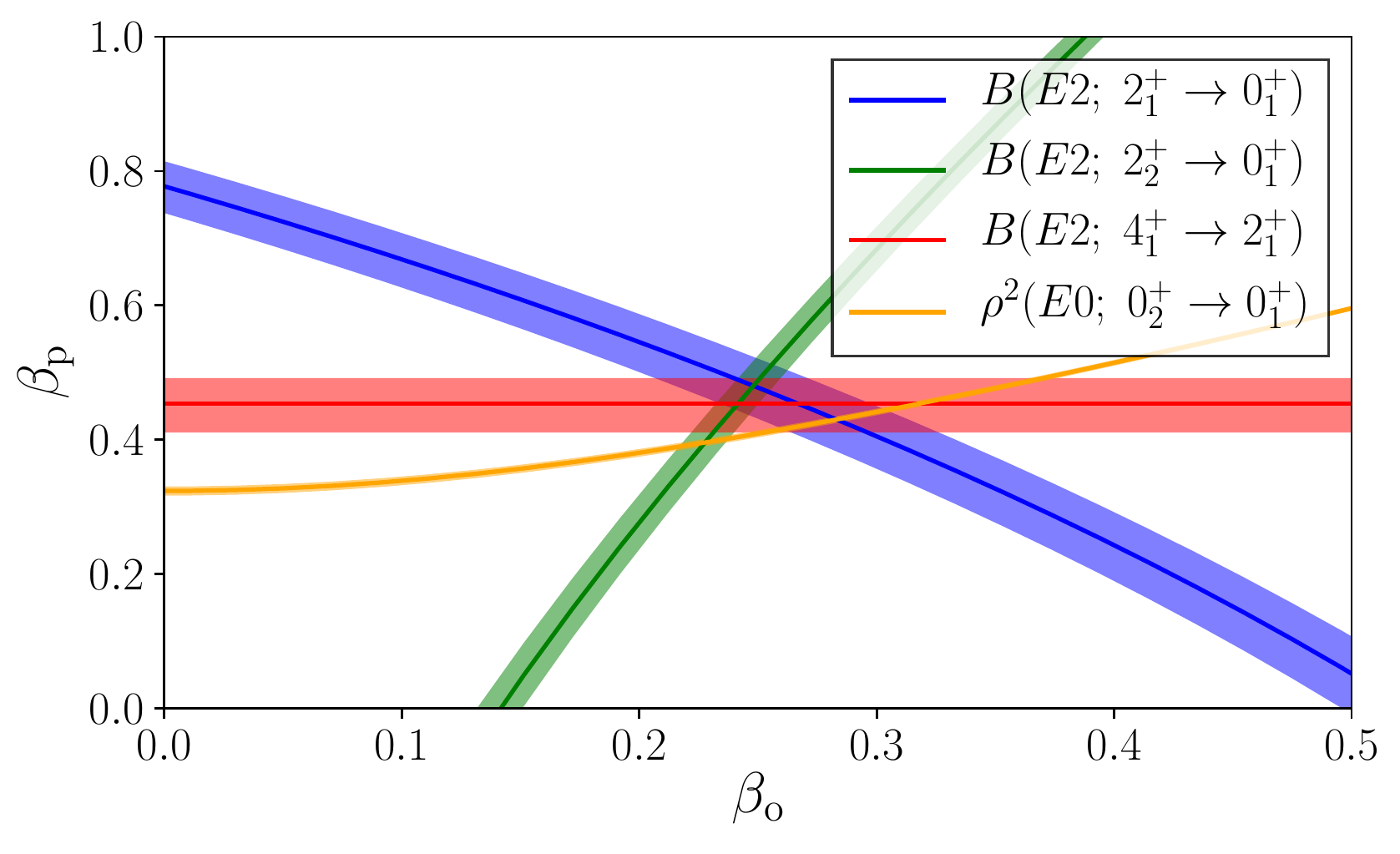}
\caption{Relation of the intrinsic deformation parameters $\beta_\text{o}$ and $\beta_\text{p}$ in the two-band mixing model. The constraints by the experimental \bevs\ for the transitions from the $2^+$ states (from this work), the $B(E2;\;4^+_1 \rightarrow 2^+_1)$~\cite{iwasaki14}, and $\rho^2(E0;\;0^+_2 \rightarrow 0^+_1)$~\cite{bouchez03}, overlap in the region around $\beta_\text{o}=0.24$ and $\beta_\text{p}=0.45$. The width of the bands represent the $1\sigma$ experimental uncertainty.}
\label{fig:betamix}
\end{figure}

The mixing model can be extended using the reduced transition probabilities $B(E2)$ and $\rho^2(E0)$. Assuming that transitions between the pure configurations are forbidden, i.e. $\langle J_\text{o}||E\lambda||J^\prime_\text{p}\rangle = 0$, the matrix elements between the physical states can be expressed as
\begin{eqnarray}
  \langle 2^+_1 || E2 || 0^+_1 \rangle & =& b_0 b_2 \langle 2^+_\text{o} || E2 || 0^+_\text{o} \rangle + a_0 a_2 \langle 2^+_\text{p} || E2 || 0^+_\text{p} \rangle\nonumber \\
  \langle 2^+_2 || E2 || 0^+_1 \rangle & =& b_0 a_2 \langle 2^+_\text{o} || E2 || 0^+_\text{o} \rangle - a_0 b_2 \langle 2^+_\text{p} || E2 || 0^+_\text{p} \rangle\nonumber \\
  \langle 4^+_1 || E2 || 2^+_1 \rangle & =& b_2 b_4 \langle 4^+_\text{o} || E2 || 2^+_\text{o} \rangle + a_2 a_4 \langle 4^+_\text{p} || E2 || 2^+_\text{p} \rangle\nonumber \\
  \langle 0^+_2 || E0 || 0^+_1 \rangle & =& a_0 b_0 (\langle 0^+_\text{o} || E0 || 0^+_\text{o} \rangle - \langle 0^+_\text{p} || E0 || 0^+_\text{p} \rangle).  
  \end{eqnarray}
The matrix elements are related to the intrinsic quadrupole moments with pure oblate or prolate deformation and thus the deformation parameters $\beta_\text{o}$ and $\beta_\text{p}$~\cite{raman01}:.
  \begin{eqnarray}
    B(E2;\;J_i \rightarrow J_f) & = & \frac{5}{16\pi} \left( eQ_0 \right)^2 \left|\langle J_i K_i 2 0 | J_f K_f\rangle \right|^2  \nonumber\\
    Q^\text{o/p}_0 &=& ZR^2 \frac{3}{\sqrt{5\pi}}\left(\beta_\text{o/p} + \frac{2}{7} \sqrt{\frac{5}{\pi}} \beta_\text{o/p}^2\right)
\end{eqnarray}
  The $4^+_1$ state is thought to be prolate deformed~\cite{iwasaki14}, and the configuration as deduced from the perturbation of the rotational energies is almost pure. It is thus reasonable to assume $a_4=1, b_4=0$. This leads to a set of four equations for the four unknowns $\beta_\text{o}$, $\beta_\text{p}$, $a_0$, $a_2$.
\clearpage
  \begin{strip}
    \begin{eqnarray}
      B(E2;\;2^+_1 \rightarrow 0^+_1) & = & \left( \frac{3}{4\pi} ZeR^2 \right)^2 |\langle2020|00 \rangle |^2 \left[ b_0 b_2 (1+0.36\beta_\text{o})\beta_\text{o} + a_0 a_2 (1+0.36\beta_\text{p})\beta_\text{p}\right]^2\nonumber \\
      B(E2;\;2^+_2 \rightarrow 0^+_1) & = & \left( \frac{3}{4\pi} ZeR^2 \right)^2 |\langle2020|00 \rangle |^2 \left[ b_0 a_2 (1+0.36\beta_\text{o})\beta_\text{o} - a_0 b_2 (1+0.36\beta_\text{p})\beta_\text{p}\right]^2\nonumber \\
      B(E2;\;4^+_1 \rightarrow 2^+_1) & = & \left( \frac{3}{4\pi} ZeR^2 \right)^2 |\langle4020|20 \rangle |^2 \left[ a_2 (1+0.36\beta_\text{p})\beta_\text{p}\right]^2\nonumber \\
      \rho^2(E0;\;0^+_2 \rightarrow 0^+_1) & = & \left( \frac{3}{4\pi} Ze\right)^2 a_0^2 b_0^2 \left( \beta_\text{o} - \beta_\text{p}\right)^2.
  \end{eqnarray}
\end{strip}
The solution to this set of equations still has several ambiguities due to the different possible signs. Using the mixing amplitudes derived above, $a_0^2 = 0.119, a_2^2= 0.744$, quartic equations connect $\beta_ \text{o}$ and $\beta_ \text{p}$. The solutions of these equations are represented by the bands in Fig.~\ref{fig:betamix}.
The $1\sigma$ experimental uncertainty bands intersect at around $\beta_\text{o}=0.24$ and $\beta_\text{p}=0.45$. The good agreement and overlap gives further confidence in the validity of the simple mixing model and suggests that no other close-lying, yet unknown states do play a major role and mix with the two bands.

\subsection{Beyond mean-field calculations}\label{sec:hfb}
Beyond mean-field calculations allow for the prediction of many nuclear properties, including spectroscopic information, for a wide range of nuclei based on the same principles.
Two types of beyond mean field calculations were used for comparison with the experimental data. Both of them use the Gogny D1S effective interaction and constraints in the mass quadrupole operators $\hat{Q}_{20}$ and $\hat{Q}_{22}$, thus spanning the full $\beta-\gamma$ plane. They do, however, vary in the methods applied for the configuration mixing. 

In the HFB-5DCH method~\cite{delaroche10} the HFB calculations are mapped on a five-dimensional collective quadrupole Hamiltonian (HFB-5DCH), which can be extracted from a microscopic  generator coordinate method (GCM) assuming that the overlaps between the different HFB states have a Gaussian form. In this way the couplings of the three rotational and two vibrational degrees of freedom are considered simultaneously without any restrictions. 
In the symmetry conserving configuration-mixing (SCCM) method~\cite{rodriguez14}, the different many-body states are calculated by mixing particle-number and angular-momentum-restored intrinsic Hartree-Fock-Bogoliubov-type wave functions (HFB), which have different (axial and non-axial) quadrupole shapes.
The intrinsic HFB states are found with the variation after particle number projection method which is more suitable to include pairing correlations than the plain HFB~\cite{anguiano01}. Finally, the mixing is performed within the GCM computing the overlaps between the different intrinsic shapes in an exact manner.

\begin{figure}[h]
\centering
\includegraphics[width=\columnwidth]{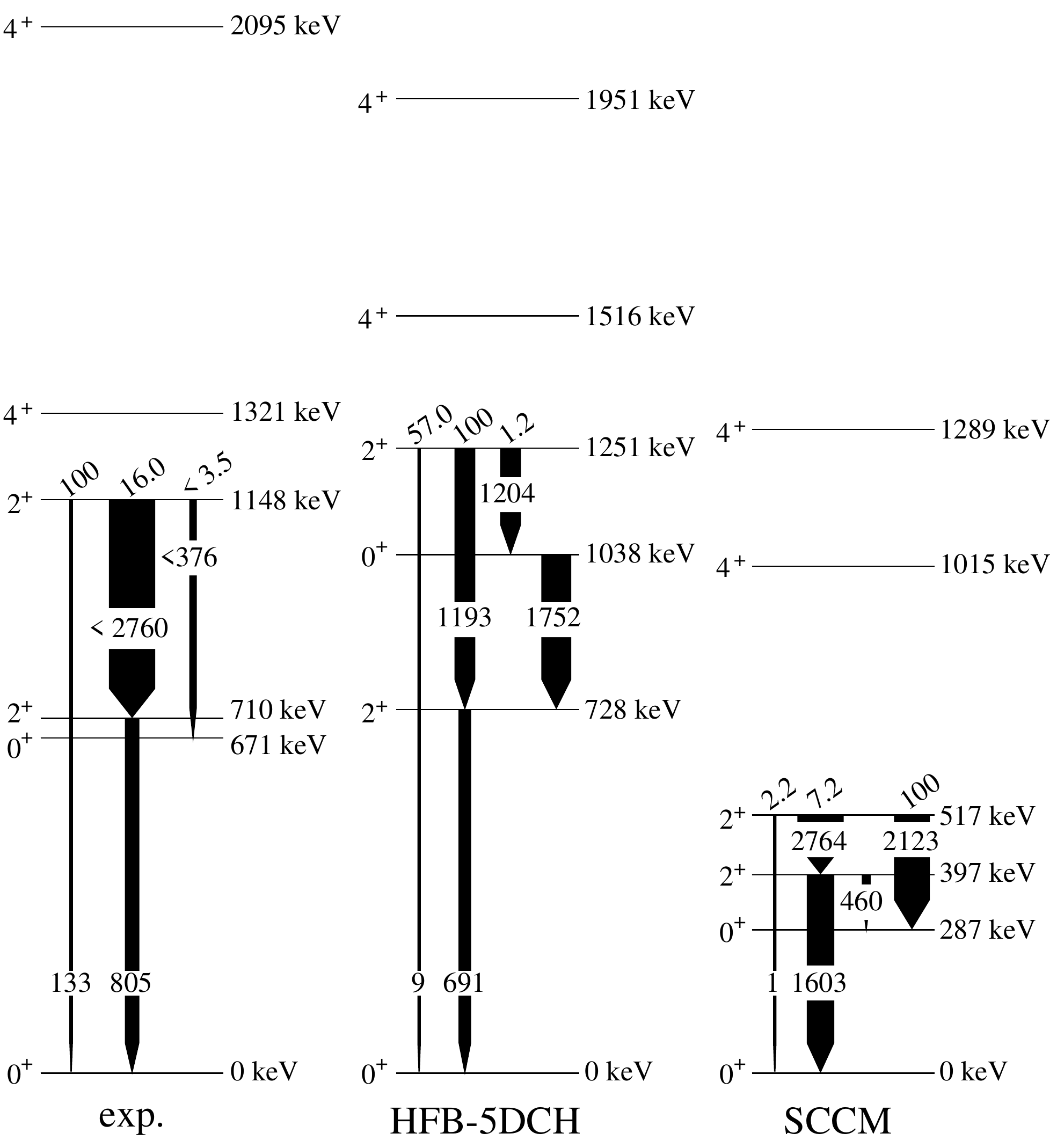}
\caption{Experimental excitation energies in keV, \bevs\ in e$^2$fm$^4$, and branching ratios in comparison to theoretical calculations using the Gogny D1S interaction in the HFB-5DCH~\cite{delaroche10} and SCCM~\cite{rodriguez14} approaches. The width of the arrows indicates the magnitude of the \bevs. For the decay of the $2^+_2$ state a pure $E2$ transition to the $2^+_1$ state has been assumed to determine an upper level for the \bev. For the decay branch to the $0^+_2$ state experimentally only an upper limit could be determined.}
\label{fig:be2_level}
\end{figure}

The HFB-5DCH calculations already showed very good agreement with experimental data for the energies and \bevs\ of \nuc{74,76}{Kr}~\cite{clement07}. In view of the softness of the potential energy surface for \nuc{72}{Kr} and a pronounced minimum at $\beta\approx0.6$~\cite{girod09}, the model space has been enlarged compared to the earlier calculations up to a deformation of $\beta=1.2$. For both types of calculation the model space has  been expanded to 11 major spherical harmonic oscillator shells. The calculated level schemes are compared to the experimental results in Fig.~\ref{fig:be2_level}.
The agreement for the excitation energies of the $2^+$ and $4^+$ states calculated in the HFB-5DCH approach with the experimental data is very good.
However, the $0^+_2$ state is predicted significantly higher (at 1038~keV) than experimentally observed. The $B(E2;\;2_1^+ \rightarrow 0_1^+)$ value is well reproduced by the calculation, for the transitions from the $2^+_2$ state the experimental values are larger, indicating a more pronounced mixing than theoretically predicted.
The quadrupole moments for the first two $2^+$ states are predicted to be $Q=11$ and $-65$~efm$^2$, indicating moderately oblate and stronger prolate deformation, respectively. 
The first $4^+$ state, calculated at 1516~keV, has a negative quadrupole moment $Q=-69$~efm$^2$, indicating prolate deformation in agreement with the shape change along the yrast states discussed in the previous section.
\begin{figure*}[t!]
  \includegraphics[width=\textwidth]{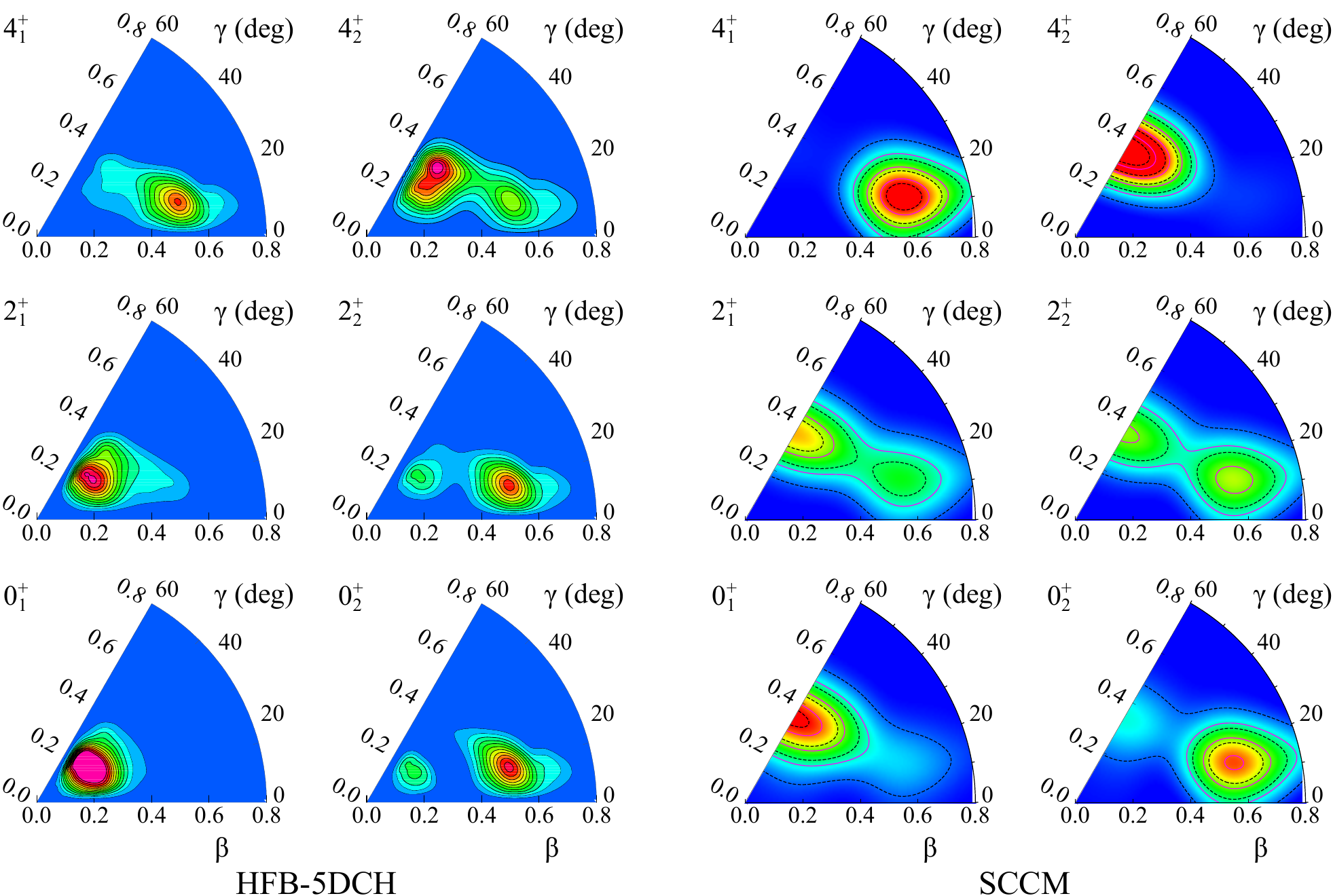}%
\caption{Probability densities calculated with the HFB-5DCH method (left) \cite{delaroche10} and collective wave functions calculated with the SCCM method (right)~\cite{rodriguez10} 
for the first and second 0$^+$, 2$^+$ and 4$^+$ states in \nuc{72}{Kr} (see text for details).}
\label{fig:collwf}
\end{figure*}

The SCCM calculation underpredicts the excitation energies in \nuc{72}{Kr}. In this case a much better agreement is observed for the heavier isotones~\cite{rodriguez14}. This could be related to an overestimation of the deformation both in the prolate and oblate bands. In fact, also the $B(E2)$ values are predicted to be larger than the experimental ones for the $2^{+}_{1}\rightarrow0^{+}_{1}$ and $2^{+}_{2}\rightarrow0^{+}_{2}$ transitions. Finally, the quadrupole moments of the $2^{+}_{1}$ and $2^{+}_{2}$ states are $-2$ and $-39$~efm$^{2}$, smaller than in the previous SCCM calculations with 9 harmonic oscillator shells (i.e.~66 and $-105$ efm$^{2}$, respectively)~\cite{rodriguez14}. As will be discussed below, this is a consequence of the strong mixing of the oblate and triaxial-prolate configurations at $J=2$ that was not present in the calculation with a smaller number of oscillator shells.

In Fig.~\ref{fig:collwf}, we compare the probability densities calculated with the HFB-5DCH method (left) with the collective wave functions calculated with the SCCM (right) approach for the two lowest bands in \nuc{72}{Kr}. In the 5DCH method these are actually provided by the wave functions (including the metric) obtained after solving the collective Hamiltonian~\cite{delaroche10}. In the SCCM case, the collective wave functions represent the weights of each intrinsic HFB state in the building of each individual GCM state~\cite{rodriguez14}. Nevertheless, both definitions can serve as a guidance to interpret the nuclear states in the intrinsic reference frame.

Both calculations show a clear picture of shape coexistence: 
In the HFB-5DCH  calculations the ground state and the first $2^+$ state are predominantly oblate deformed, with a shape change towards prolate deformation at $J=4$. The yrare $0^+$ and $2^+$ states are predominantly prolate deformed, but with an increasing oblate deformed component, which becomes dominant in the $4^+_2$ state. A clear picture of a shape inversion for the yrast band emerges, in agreement with the conclusion drawn from the experimental excitation energies, but later between $J= 2$ and 4.

The SCCM approach also predicts \nuc{72}{Kr} to exhibit shape coexistence, but with weak mixing of the intrinsic configurations of the ground (mostly oblate) and first excited (triaxial-prolate) $0^+$ states. The $2^+$ states, on the other hand, show strong mixing between these configurations reproducing the experimental trend. The shape inversion occurs at $J=4$, where the mixing disappears, the triaxial-prolate band becomes yrast and the oblate band is at a higher excitation energy. 
These results are slightly different to the ones  reported previously~\cite{rodriguez14}. These earlier calculations were performed with only nine oscillator shells, in view of the large computational burden for calculating long isotopic chains. Both the excitation energies and the shape mixing were not fully converged for this previous model space. 

Although the present SCCM calculations show a qualitative agreement with the experimental results (shape coexistence observed in the first $0^+$ states, shape mixing in the $2^+$ states), some quantitative discrepancies are found (see Fig.~\ref{fig:be2_level}), i.e.~the excitation energies are too small and the in-band \bevs\ are too large. These deficiencies could be related to two effects. On the one hand, proton-neutron pairing correlations are missing and are expected to be large in $N=Z$ systems. Such correlations could push down the ground-state energy, resulting in larger excitation energies for the $0^+_2$ and $2^+$ states. On the other hand, the present SCCM calculations tend to overestimate the deformation due to the use of a nuclear interaction with the D1S parametrization.  
However, fitting the Gogny force including SCCM techniques, albeit desirable, is beyond the scope of the present study. 

\section{Summary}
The $N=Z$ nucleus \nuc{72}{Kr} has been studied by inelastic scattering. Four new transitions have been observed for the first time in the scattering off a low-Z Be target and placed in the level scheme using $\gamma$-ray coincidences. The new $2^+_2$ state at 1148~keV has also been observed in the scattering off a high-Z Au target. The consistent analysis of both data sets allowed for the extraction of nuclear deformation parameters and \bevs\ for the $2^+$ states. The value for the $2^+_1$ state, $B(E2; 0^+_1\rightarrow 2^+_1)\allowbreak=\allowbreak4023(81)_\text{stat.}(290)_\text{syst.}(380)_\text{theo.}$~e$^2$fm$^4$, is in agreement with previous measurements~\cite{gade05,iwasaki14}, while $B(E2; 0^+_1\rightarrow 2^+_2)\allowbreak=\allowbreak665(39)_\text{stat}(58)_\text{syst.}(63)_\text{theo.}$~e$^2$fm$^4$ was obtained for the first time in the present experiment. The analysis in a two-band mixing model corroborates the interpretation of shape coexisting states, with an oblate-deformed ground state and a prolate deformed $0^+$ shape isomer and an inversion of the shape of the yrast states starting already with the $2^+$ states.
HFB-5DCH calculations with the Gogny D1S interaction predict an oblate-prolate shape-inversion scenario and describe the measured B(E2) values reasonably well. 
Calculations using the SCCM method, also predict a shape coexistence of less oblate and more triaxial-prolate deformed configurations. A strong shape mixing is found for the first two $2^+$ states. However, the current SCCM method predicts larger \bevs\ and smaller excitation energies than the experimental values due to an overestimation of the deformation.

\begin{acknowledgement}
We would like to thank the RIKEN accelerator and BigRIPS teams for providing the high intensity beams.
We thank T. Furumoto for providing us with the optical potentials and A. Moro for giving access to an unpublished version of the FRESCO code. 
This work has been supported by UK STFC under grant numbers ST/L005727/1 and ST/P003885/1, the Spanish Ministerio de Econom\'ia y Competitividad under grants FPA2011-24553, FPA2014-52823-C2-1-P, and PGC2018-094583-B-I00, the Program Severo Ochoa (SEV-2014-0398), the European Research Council through the ERC Grant No. MINOS-258567, NKFIH (NN128072), and by the \'UNKP-19-4-DE-65 New National Excellence Program of the Ministry of Human Capacities of Hungary. G. K. acknowledges support from the J\'anos Bolyai research fellowship of the Hungarian Academy of Sciences. K. W. acknowledges the support from the Spanish Ministerio de Econom\'ia y Competitividad RYC-2017-22007.
\end{acknowledgement}
\bibliographystyle{spphys}
\bibliography{kr72resubmitted}

\end{document}